\documentclass[aps, 10pt, a4paper, twocolumn,floatfix, showkeys, superscriptaddress]{revtex4-2}

\usepackage[utf8]{inputenc}
\usepackage{ragged2e}
\usepackage{array}
\usepackage{graphicx}
\usepackage{mathtools}
\usepackage{amsfonts, amssymb}
\usepackage{calrsfs}
\usepackage{silence}
\usepackage[most]{tcolorbox}

\usepackage{mlmodern}
\usepackage{siunitx}

\usepackage[hidelinks]{hyperref}

\newcolumntype{L}[1]{>{\raggedright\arraybackslash}p{#1}}

\setcitestyle{numbers,square,comma}

\makeatletter
\def\NAT@sort{\@ne}
\def\NAT@cmprs{\@ne}
\makeatother

\begin{document}

\title{Large-Language Models as a Cognitive Virus}


\providecommand{\CSL}{Complex Systems Lab, Universitat Pompeu Fabra (MELIS), Dr. Aiguader 88, 08003 Barcelona, Spain}
\providecommand{\ICREA}{Instituci{\'o} Catalana de la Recerca i Estudis Avancats (ICREA), Passeig Llu\'is Companys 23, 08010 Barcelona, Spain.}
\providecommand{\IBE}{Institut de Biologia Evolutiva, CSIC-UPF, Passeig Mar{\'i}tim de la Barceloneta 37, 08003 Barcelona, Spain.}
\providecommand{\SFI}{Santa Fe Institute, 1399 Hyde Park Road, Santa Fe NM, United States.}
\providecommand{\BCOM}{Barcelona Computational Foundation (BCOM) and Neuroelectrics, Barcelona, Spain}
\providecommand{\SYSBIO}{Instituto de Biología Integrativa de Sistemas (I2SysBio), CSIC-Universitat de València, Paterna, València, 46980, Spain}
\providecommand{\TUFTS}{Allen Discovery Center, Tufts University, Medford MA, United States.}
\providecommand{\WYSS}{Wyss Institute for Biologically Inspired Engineering, Harvard University, Boston MA, United States.}
\providecommand{\OIST}{Okinawa Institute of Science and Technology Graduate University, Onna, 904-0495, Japan. }
\providecommand{\CMNL}{Complex Multilayer Networks Lab, Department of Physics and Astronomy 'Galileo Galilei', University of Padua, Via Marzolo 8, 35131 Padova, Italy} 
\providecommand{\INFN}{Istituto Nazionale di Fisica Nucleare, Sez. Padova, 35131 Padova, Italy} 
\providecommand{\PCNM}{Padua Center for Network Medicine, Via Marzolo 8, 35131 Padova, Italy} 
\providecommand{\PNC}{Padua Neuroscience Center, Via Giuseppe Orus, 2, Italy}

\author{Ricard Sol\'{e}}
\email[Corresponding author, ]{ricard.sole@upf.edu}
\affiliation{\CSL}
\affiliation{\ICREA}
\affiliation{\IBE}
\affiliation{\SFI}

\author{Giulio Ruffini}
\email[Corresponding author, ]{giulio.ruffini@bcom.one}
\affiliation{\BCOM}

\author{Francesca Castaldo}
\email[Corresponding author, ]{francesca.castaldo@bcom.one}
\affiliation{\BCOM}

\author{Marco Tuccio}
\affiliation{\CSL}
\affiliation{\IBE}

\author{Luis F. Seoane}
\affiliation{\IBE}
\affiliation{\OIST}

\author{Manlio de Domenico}
\affiliation{\CMNL}
\affiliation{\INFN}
\affiliation{\PCNM}
\affiliation{\PNC}

\author{Santiago F. Elena}
\affiliation{\SYSBIO}
\affiliation{\SFI}

\author{David C. Krakauer}
\affiliation{\SFI}

\author{Michael Levin}
\affiliation{\TUFTS}
\affiliation{\WYSS}

\vspace{0.4 cm}
\begin{abstract}
\vspace{0.2 cm}
Large-language models (LLMs) are rapidly becoming part of human culture, reshaping how information is produced, transmitted, and used. Here we propose that their diffusion can be understood through a viral analogy, with LLM use spreading through populations, becoming embedded in cognitive and cultural practices. We model transitions among uncoupled, coupled, and persistently dependent users, and show that the interplay between social transmission, recovery, and collective reinforcement can generate tipping points and technological lock-in. A central consequence is the possibility of runaway dynamics: once a critical threshold is crossed, small increases in adoption can trigger rapid population-level shifts toward persistent dependence, with abrupt losses in cognitive competence. The same framework, however, identifies conditions for cognitive immunization, based on reducing transmission and facilitating reversibility. Our results highlight how LLM adoption may involve nonlinear collective transitions with important consequences for cognitive autonomy.
\end{abstract}


\maketitle

\begin{tcolorbox}[ colback=blue!10, colframe=black!100, boxrule=0.8pt, arc=1mm, left=6pt, right=6pt, top=5pt, bottom=5pt ] \textbf{Significance Statement.} 
Large language models are rapidly becoming embedded in everyday cognitive and cultural practices, yet we still lack simple theoretical frameworks for understanding how their adoption may reshape collective patterns of cognition. Here we introduce the idea of LLMs as cognitive viruses: culturally transmitted technologies whose spread is promoted by their usefulness but can also increase dependence through cognitive offloading. A dynamical model shows that social facilitation can generate runaway transitions from predominantly autonomous cognition to widespread technological dependence. These results suggest that abrupt collective changes can emerge even when individual adoption is gradual. By connecting epidemic dynamics, cultural transmission, and cognitive offloading, our framework identifies general mechanisms that may govern human–AI coevolution and provides a basis for ``cognitive immunization''.
\end{tcolorbox}

\section{Introduction}

The emergence of human language is widely regarded as a major evolutionary 
transition \cite{szathmary1995major}.  While it shares key features with genetic transmission \cite{nowak1999evolution}, language established a new system of inheritance and collective memory. It allows information to accumulate, be passed on, and recombine 
over generations, thus supporting cumulative culture 
\cite{tomasello1999cultural,boydricherson1985culture}. In this sense, languages 
can be viewed as population-level cultural systems with partially autonomous 
dynamics: they diversify, compete, hybridize, spread, and sometimes become 
extinct, showing patterns reminiscent of species embedded in ecological 
communities \cite{sole2010diversity}. Their persistence depends largely on social 
learning, through which individuals acquire words, grammatical constructions, 
meanings, and communicative conventions by observing and interacting with 
others, especially during childhood \cite{tomasello2003constructing}. Because 
linguistic structures propagate by entering developing minds, relying on their 
exceptional plasticity, and recruiting their speakers as new vectors of 
transmission, language has sometimes been described as a viral entity 
\cite{burroughs1970electronic,aoki1999language}. 

The metaphor acquired a concrete technological counterpart with the emergence 
of computer viruses: self-propagating informational structures capable of 
entering a host system, modifying its operation, and recruiting it for further 
transmission \cite{cohen1987computer}. In parallel, 
theories of cultural evolution formalized the idea that socially transmitted information can exhibit 
variation, differential persistence, and inheritance, whether framed through 
gene-culture coevolution \cite{feldman1996gene,gintis2011gene,whitehead2019reach} or through the concept of memes \cite{dawkins1976selfish,blackmore2000meme,dennett2017bacteria}. 
Together, these traditions point towards a broader class of entities, the ``viruses of the mind'' \cite{dawkins1993viruses,brodie2009virus} whose evolution depends not on a particular material substrate, but on their capacity to reproduce and persist across biological, cognitive, or computational hosts.

In language, what propagates virally are patterns---from words to ideas---that reproduce through imitation, mutate through use, and persist through brains, media, institutions, and technologies. This points to cognition as distributed across minds and external structures---the extended mind \cite{clark1998extended}---which do not merely store information but also reshape cognitive tasks \cite{norman1991cognitive,hutchins1995cockpit}. Such artifacts can be \textit{complementary}, strengthening capacities beyond their immediate use, or \textit{competitive}, improving performance while potentially weakening the underlying skill \cite{krakauer2016will,krakauer2026competitive,fasoli2018substitutive}. Language itself is a two-sided cognitive technology: it externalizes, stabilizes, and recombines thought \cite{Clark1998MagicWords}, while linguistic categories and narratives reshape perception and reasoning \cite{WinawerEtAl2007,LupyanWard2013,FauseyEtAl2010,ThibodeauBoroditsky2011}. Through cultural transmission, language is also adapted to the cognitive and communicative demands of its users \cite{KirbyEtAl2008,christiansen2016creating,GelmanRoberts2017}, so that its propagation both extends cognition and transforms the minds that depend on it.

This reciprocal transformation also characterizes communication technologies. The Internet and instant messaging blurred distinctions between speech and writing \cite{Crystal2006Internet,Baron2008AlwaysOn}, while mobile phones and smartphones encouraged abbreviated, compressed forms of expression that largely reflect linguistic flexibility rather than declining literacy \cite{LingBaron2007,LyddyEtAl2014,RuizTrilloEtAl2026}. Social media further reorganized the structure and social selection of communication \cite{Boyd2010NetworkedPublics,Androutsopoulos2014ContextCollapse}, integrating text, hyperlinks, multimodal expression, and algorithmically selected content \cite{Kress2010Multimodality,Zappavigna2012DiscourseTwitter}. Such technologies do not merely transmit language: their affordances reshape how it is produced, circulated, interpreted, and selected.

LLMs represent a further transition. Unlike previous media, they do not merely constrain the form or circulation of human expression but actively participate in its production, reformulation, and evaluation. The relevant question, therefore, is not simply whether LLMs alter language use, but whether they reorganize the coupled system formed by language, cognition, and technology. This comparison may help identify which cognitive capacities these systems amplify, which they displace, and under what conditions sustained reliance may produce cognitive reorganization or erosion. In this respect, LLMs extend a feature shared, to varying degrees, by earlier information technologies: cognitive offloading, understood as the delegation of cognitive operations to external tools and representations.

Two technological precursors bridge natural language and the LLM era: computer viruses (CVs) \cite{cohen1989computational,cass2001anatomy,parikka2007digital} and programming languages (PLs) \cite{sammet1972programming}. CVs reproduced key features of biological viruses: compact informational structures that exploit host machinery and evolve to evade detection. Their emergence triggered an arms race with antivirus software \cite{nachenberg1997computer}, influenced artificial life \cite{spafford1994computer} and network science \cite{pastor2001epidemic,lloyd2001viruses,keeling2005networks}, and fostered modern cybersecurity and computer immunology \cite{forrest1997computer,dasgupta2008immunological}. PLs spread across computers, institutions, and communities \cite{valverde2015punctuated,valverde2015cultural}, becoming an operational substrate of the digital revolution and, alongside neural networks, paving the way for artificial intelligence \cite{hey2015computing}. Both depend on technological hosts and reshape their environments. LLMs combine aspects of each: like PLs, they provide a new human--machine interface; like CVs, they propagate through users and digital ecosystems. Yet their capacity for cognitive offloading is unprecedented \cite{gerlich2025ai}.

Here, we address our previous question by treating LLMs as engines of cognitive and social change whose rapid diffusion is transforming human capacities. We first compare LLMs with infectious agents and then develop an explicit population model of their propagation, drawing on mathematical approaches from epidemiology \cite{anderson-may} that have also been applied to technological adoption \cite{karshenas1993rank,dahlke2024epidemic}. The model describes transitions among host-coupling states, which we link to an illustrative measure of cognitive competence to distinguish the consequences of different coupling regimes from the underlying bifurcation structure and explore potential interventions. It does not identify any single entity as the viral analogue: LLM ecosystems contain culturally transmitted practices and content alongside technologically evolving model lineages, and these need not coincide. Instead, the analogy concerns a broader feedback loop in which persistent technological lineages are instantiated in external machinery, modify their human host environment, and thereby influence their own propagation. This homology is already partly present in conventional and adaptive software \cite{lehman1984program,lehman1996feedback,fitzgerald2017continuous,ros2024face}, but LLMs deepen it by becoming integrated into cognitive processing itself.

\begin{figure*}[htbp]
\centering
\includegraphics[width=16 cm]{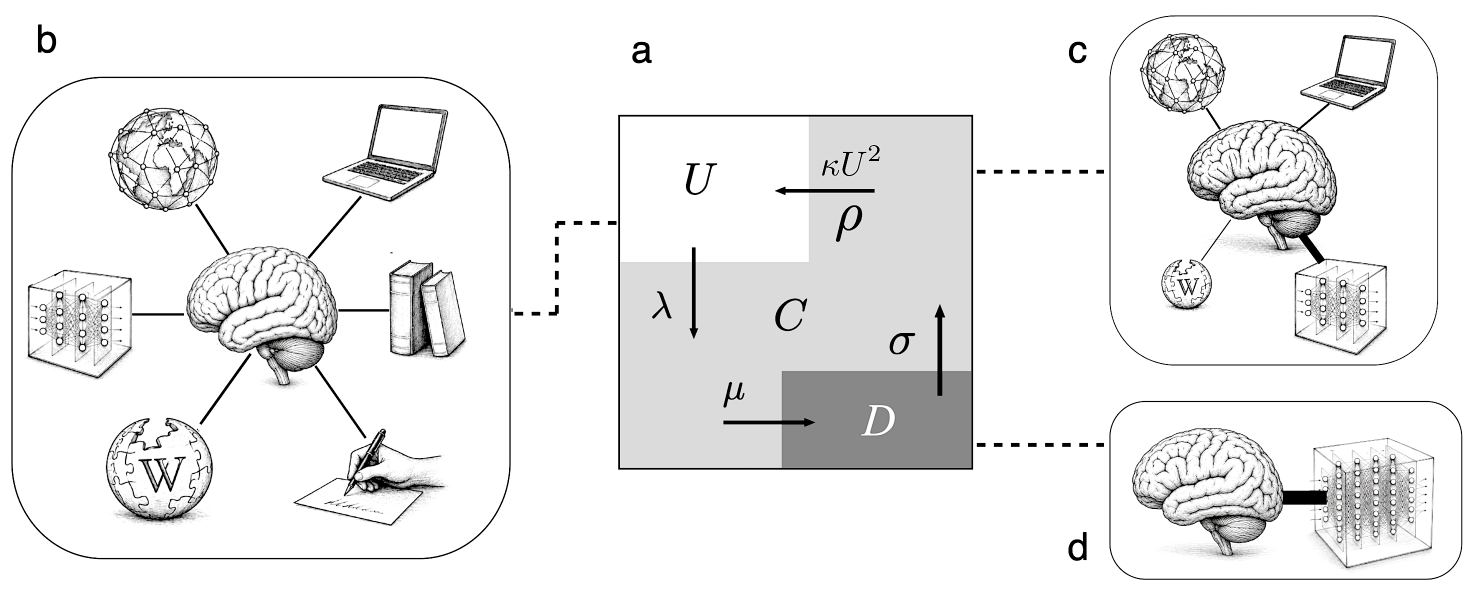}
\caption{\footnotesize
Population-level host-state model within an LLM propagation ecology.
(a) Individuals move among three states: uncoupled or weakly coupled users
$U$; autonomous coupled users $C$, who use LLMs while retaining reading,
writing, reasoning, verification, and access to alternative information
sources; and persistently dependent users $D$, for whom LLM-mediated
cognitive operations have become strongly substitutive.
Exposure to socially,
institutionally, or platform-mediated LLM practices drives the transition
$U\rightarrow C$ at rate $\lambda$, whereas abandonment returns users to
the uncoupled state at rate $\rho$. Regular use develops into dependency
at rate $\mu$, while training, verification practices, or deliberate
cognitive friction restore autonomous use at rate $\sigma$. 
(b) In the uncoupled state, cognition is distributed across the brain and
a heterogeneous external information environment, including books,
writing, computers, Wikipedia, and the Internet.
(c) In the coupled but autonomous state, the LLM becomes an additional
component of this cognitive ecology, while other information sources and
independent cognitive operations are retained.
(d) In the dependent state, interaction with the LLM becomes dominant,
progressively replacing alternative cognitive supports and concentrating
cognitive activity within a single human-machine coupling. 
The model therefore describes transitions among human host/coupling states
associated with LLM use. It represents one population-level layer of the
broader viral ecology and does not by itself model reproduction of the
model/program lineage.
}
\label{fig:threescenarios}
\end{figure*}
The viral analogy does not imply that LLM-human interactions are intrinsically parasitic. Biological viruses range from pathogens to mutualists and evolutionary partners \cite{marquez2007virus,ryan2009virolution,roossinck2011good,barr2013bacteriophage,roossinck2017symbiosis}, and the effects of LLMs likewise depend on how they are used. They can enhance exploration, access to expertise, and productivity \cite{noy2023experimental,brynjolfsson2025generative}, but can also promote cognitive offloading, dependence, and loss of competence \cite{risko2016cognitive}, while individual benefits may not scale to the population level \cite{doshi2024generative}. The analogy is therefore useful because LLM use can generate persistence and transmission loops, both through culturally propagating patterns of use and through model lineages sustained and modified by technological and economic feedbacks. The model below focuses on one part of this ecology: transitions among human states of cognitive coupling.

\section{Population dynamics of LLM-mediated cognitive coupling}

Classical epidemic models describe the transmission of biological agents through transitions between host states \cite{anderson-may}. A similar mathematical language has been used for technological diffusion, where adoption depends on exposure to previous users and social reinforcement \cite{bass1969,karshenas1993rank}, as well as in multicompartment models of drug \cite{white2007heroin,battista2019modeling,van2022review} and social media \cite{chou2005review,j2014internet,alemneh2021mathematical} addictions. 
At the level modeled here, we do not represent the reproduction or evolution
of the model/program lineage explicitly. Instead, we coarse-grain one
epidemiological layer of the larger LLM ecology: transitions among human
host/coupling states.
We consider three population states (Fig.~\ref{fig:threescenarios}a): uncoupled or weakly coupled individuals $U$, regular users $C$ who retain cognitive autonomy, and dependent users $D$ who persistently delegate cognitive operations to the model. Their dynamics are 
\begin{eqnarray} \frac{dU}{dt}&=&-\lambda UC+\rho C+\kappa U^2C,\\ 
\frac{dC}{dt}&=&\lambda UC-(\mu+\rho)C+\sigma D-\kappa U^2C,\\ 
\frac{dD}{dt}&=&\mu C-\sigma D, 
\end{eqnarray} 
with $U+C+D=1$. Here, $\lambda$ measures the effective spread of LLM practices through social and institutional exposure, $\rho$ the return from regular use to the uncoupled state, $\mu$ the transition from regular use to persistent dependency, and $\sigma$ recovery from dependency. These rates coarse-grain heterogeneous individual processes into population-level transitions rather than representing single psychological mechanisms. The nonlinear term $\kappa U^2C$ introduces a cooperative (Allee-like) mechanism \cite{courchamp2008allee}. 
 The underlying assumption is that LLM-independent cognitive practice is socially reinforced and cultural expectations that reward independent reasoning become more effective when autonomous individuals are common. The quadratic dependence on $U$ represents this positive frequency dependence, whereas the factor $C$ restricts the restoring effect to individuals engaged with LLMs. 
Accordingly, the incidence term $\lambda UC$ should be interpreted as an
effective host-side social or institutional transmission pressure. As in
biological compartment models, the fact that the state variables describe
hosts does not imply that a host state is the pathogen. The equations model
the epidemiology of coupling and therefore do not, by themselves, determine
the identity of the viral analogue.

Mean-field approximations like the one we take here provide a valuable, analytically tractable baseline for identifying the mechanisms underlying collective behavior \cite{sole2011phase}. Here, individuals sample population-level frequencies, with connectivity absorbed into the effective rate $\lambda$. However, real social and technological networks are heterogeneous and correlated~\cite{de2026decoding}: connectivity distributions can shift epidemic thresholds and immunization outcomes~\cite{newman2002spread,pastor2015epidemic,pastor2001epidemic,may2001infection,pastor2002immunization,cohen2003efficient}, while other network properties can further modify diffusion~\cite{boguna2002epidemic,newman2002assortative,newman2003properties,miller2009spread,salathe2010dynamics}. Communication topology could also shape collective behavior in LLM-agent networks (as they do in biological epidemics \cite{zomer2026unraveling}), particularly when cognitive states reshape social ties (again, as for biological pathogens \cite{gross2006epidemic}). At larger scales, interconnected regions and institutions can generate invasion thresholds and topology-dependent patterns~\cite{colizza2007invasion,brechtel2018master,nauta2024topological}. Our formulation therefore offers a useful reference for future network and spatial extensions.

The cooperative component at the population-level has strong precedents in social and ecological dynamics. The quadratic term, commonly used in evolutionary ecology models to introduce mutualisms  \cite{kefi2007local,scheffer2009early,sole2011phase} assumes that autonomous cognitive practices are frequency dependent: schools, workplaces, peer groups, and norms can make independent work easier to maintain when it remains common. Threshold models of collective behavior show theoretically that such frequency dependence naturally produces critical transitions in humans \cite{granovetter1978threshold,ball2004critical,castellano2009statistical}. In particular, it has been shown that social conventions can undergo genuine tipping \cite{centola2018tipping,ashery2025emergent} when a minority of the group was sufficient to move a population from one convention to another. 

Dependency is assumed to arise primarily through regular use, $C\rightarrow D$, rather than directly from $U$. These simplifications isolate the interaction between contagion-like technological adoption and collective protection of cognitive autonomy, allowing us to identify conditions for abrupt and potentially irreversible population-level transitions. Figure \ref{fig:threescenarios} connects these population states with different forms of human-LLM coupling. Uncoupled individuals remain embedded in a diversified cognitive ecology, while making little or no use of LLMs (Fig. \ref{fig:threescenarios}b). Regular users incorporate LLMs while retaining substantial independent cognitive activity (Fig. \ref{fig:threescenarios}c), whereas in the dependent state the LLM becomes the dominant interface for performing tasks and accessing information (Fig. \ref{fig:threescenarios}d).

Transitions among these regimes can result from social exposure, institutional adoption, repeated reliance on generated outputs, abandonment of the technology, or progressive cognitive offloading. Cognitive offloading is itself a normal and often adaptive component of human cognition \cite{risko2016cognitive,gerlich2025ai,jose2025cognitive}. Reading and writing provide a paradigmatic example: literacy recruits and reorganizes preexisting neural circuits \cite{dehaene2015illiterate}, while external symbolic structures extend memory and enable new forms of reasoning. Crucially, however, external support can either generate new internal competence or substitutes for it. 

Recent work on LLM-assisted writing suggests that increased externally supported performance can be accompanied by reduced cognitive engagement, recall, or sense of authorship \cite{kosmyna2025your}. It is therefore useful to distinguish between \emph{scaffolding} and \emph{substitution}. Scaffolding reduces immediate cognitive demands while preserving or increasing the user's subsequent capacity to perform the task; substitution removes the need to perform the underlying cognitive operation. Following \cite{krakauer2016will}, these correspond broadly to \emph{complementary} and \emph{competitive cognitive artifacts}. An LLM may thus operate as a tutor whose arguments are reconstructed, challenged, and verified, or as a substitute that performs synthesis, evaluation, and composition on behalf of the user. The relevant distinction is therefore not simply between using and avoiding LLMs, but between forms of coupling and, ultimately, what cognitive capacities remain when the tool is withdrawn. The transition $C\rightarrow D$ provides a minimal representation of this shift from complementary use toward persistent substitutive dependency.

\begin{figure}[htbp]
    \centering
    \includegraphics[width=7 cm]{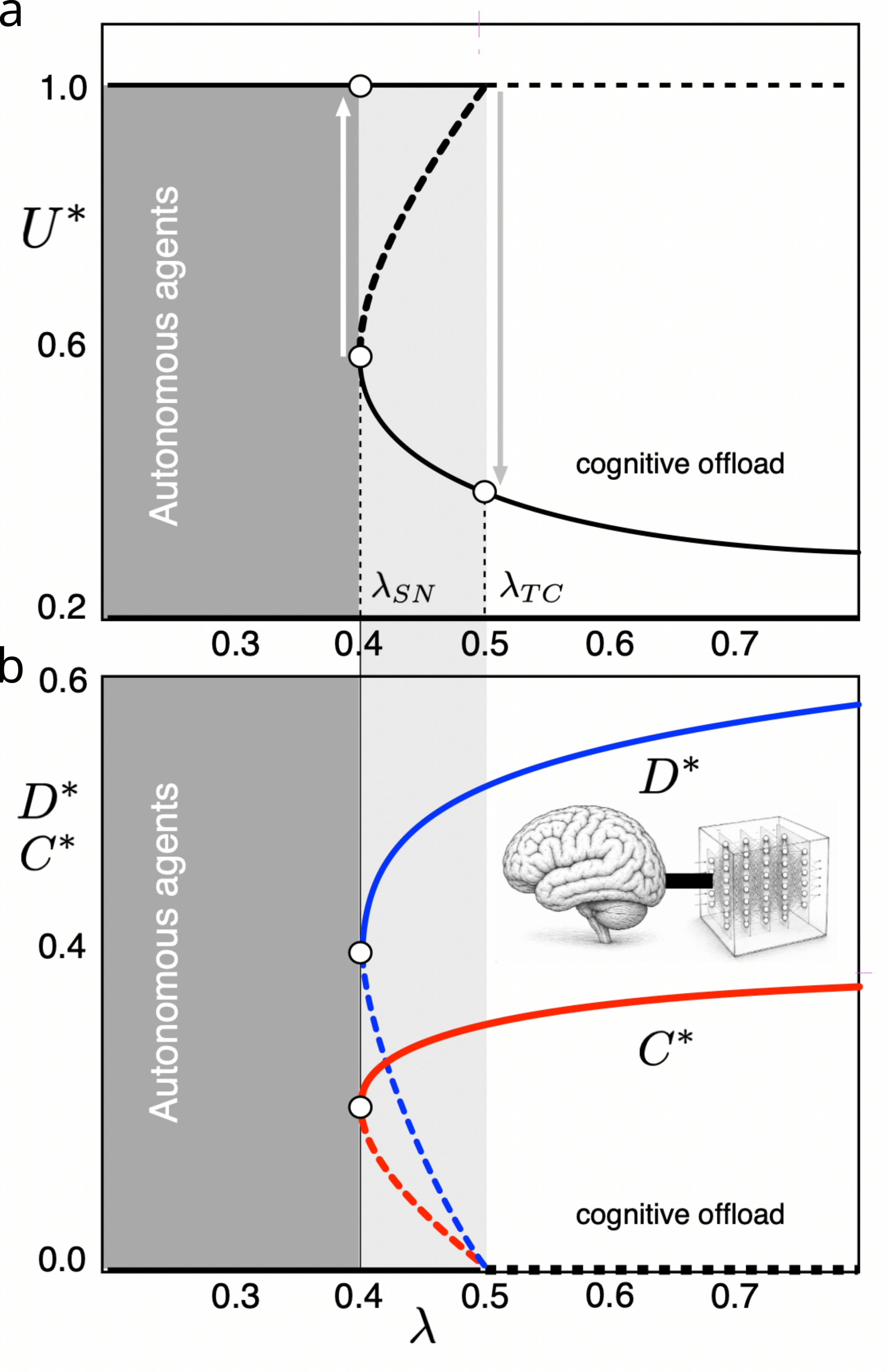}
\caption{\footnotesize
Bifurcation structure of the LLM propagation model. (a) Equilibrium fraction of uncoupled individuals, $U^*$, as a function of the transmission parameter $\lambda$. For $\lambda_{\rm SN}<\lambda<\lambda_{\rm TC}$, the system is bistable: the fully uncoupled state and a coupled state are both stable and are separated by an unstable branch (dashed). Increasing $\lambda$ drives an abrupt transition to the coupled regime at the transcritical threshold $\lambda_{\rm TC}$, whereas recovery occurs only after $\lambda$ is reduced below the saddle-node threshold $\lambda_{\rm SN}$, generating hysteresis and technological lock-in. (b) Corresponding equilibrium fractions of regular users, $C_{\pm}^*$ (red), and dependent users, $D_{\pm}^*$ (blue). Solid curves denote stable equilibria and dashed curves unstable branches. Beyond the tipping point, the sharp decrease in $U^*$ is accompanied by a rapid increase in both regular LLM use and persistent cognitive dependency, representing population-level cognitive offloading.
Parameters used: $\rho=0.10,\kappa=0.40,\mu=0.20,\sigma=0.10$.}
    \label{fig:LLMbifurcations1}
\end{figure}

The model displays qualitatively distinct long-term regimes of LLM use. Using $D=1-U-C$, the dynamics can be written as a two-dimensional system
\begin{align} 
{dU \over dt} &= C f(U),\\ 
{dC \over dt} &= -C f(U)-\mu C+\sigma(1-U-C), \end{align} 
where $f(U)=\rho-\lambda U+\kappa U^2$. An equilibrium (fixed point) is the fully uncoupled state $E_U=(1,0,0)$, corresponding to a population in which LLM-mediated cognitive coupling is absent. The other points satisfy $f(U^*)=0$ and $D^*=(\mu/\sigma)C^*$, giving 
\begin{equation} 
C_{\pm}^*=\frac{\sigma}{\mu+\sigma}(1-U_{\pm}^*), \qquad D_{\pm}^*=\frac{\mu}{\mu+\sigma}(1-U_{\pm}^*), \label{eq:CDstar} 
\end{equation} 
where $U_{\pm}^*$ are the two roots defined as: 
\begin{equation} 
U_{\pm}^*= \frac{\lambda\pm\sqrt{\lambda^2-4\kappa\rho}} {2\kappa}. \label{eq:Upm_results} 
\end{equation}
The two branches exist when $\lambda\geq\lambda_{\rm SN}=2\sqrt{\kappa\rho}$. The uncoupled equilibrium remains stable while 
\begin{equation} \lambda<\lambda_{\rm TC}=\rho+\kappa 
\end{equation} 
and loses stability at $\lambda_{\rm TC}$ through a transcritical bifurcation. When $\kappa>\rho$, the saddle-node occurs before this loss of stability, producing the bistable interval 
\begin{equation} 
2\sqrt{\kappa\rho}<\lambda<\rho+\kappa. \label{eq:bistable_interval} \end{equation} 
Within this interval, both the uncoupled and the coupled state coexist, separated by the unstable branch (dashed curves in Fig. \ref{fig:LLMbifurcations1}). Consequently, increasing $\lambda$ from the uncoupled state leaves the population near $U=1$ until $\lambda_{\rm TC}$ is reached, whereas decreasing $\lambda$ from the coupled state does not restore the uncoupled regime until $\lambda_{\rm SN}$ is crossed. The difference between these thresholds, 
\begin{equation} 
  \lambda_{\rm TC}-\lambda_{\rm SN} = \left(\sqrt{\kappa}-\sqrt{\rho}\right)^2, 
  \label{eq:dLambda}
\end{equation} 
defines the width of the hysteretic region and provides a simple mechanism for technological lock-in.

These transitions are illustrated in Fig. \ref{fig:LLMbifurcations1}. Figure \ref{fig:LLMbifurcations1}a shows the equilibrium uncoupled fraction $U^*$ as $\lambda$ is varied: a stable coupled branch and a saddle emerge at $\lambda_{\rm SN}$, while the uncoupled state loses stability only at $\lambda_{\rm TC}$. The corresponding regular-use and dependent fractions are shown in Fig. \ref{fig:LLMbifurcations1}b and follow directly from Eq.~\eqref{eq:CDstar}. Figure \ref{fig:LLMbifurcations2}b summarizes the bifurcation structure in the $(\lambda,\kappa)$ plane. Genuine bistability occurs only for $\kappa>\rho$ and between the saddle-node boundary $\lambda=2\sqrt{\kappa\rho}$ and the transcritical boundary $\lambda=\rho+\kappa$. For $\kappa<\rho$, the formal saddle-node lies outside the physical simplex and the transition is instead continuous, occurring through a forward transcritical bifurcation. At $\kappa=\rho$, the two thresholds coincide at $\lambda=2\rho$, marking the boundary between continuous adoption and history-dependent discontinuous transitions.

\section{cognitive competence across the transition}

The population-level transition can have consequences for human competence,
but their sign is not fixed by the bifurcation itself. To illustrate one
possible consequence, we assign each state a relative cognitive competence, $\Gamma_u$, $\Gamma_c$, and $\Gamma_d$, and define
the average
\begin{equation}
\langle\Gamma\rangle
=\Gamma_u U+\Gamma_c C+\Gamma_d D.
\end{equation}
Here $\Gamma$ should be narrowly interpreted as cognitive competence (CC): the capacity available to the human when external support
is removed, rather than the total capability of the coupled human-AI
system. To illustrate a substitutive-use regime, we use
$\Gamma_u=1$, $\Gamma_c=0.5$, and $\Gamma_d=0.1$, corresponding to
progressively greater loss of cognitive competence with increasing
dependence. This ordering is an illustrative modelling assumption, not
a general claim about LLM use. In scaffolded or augmentative regimes,
coupling could leave subsequent CC unchanged or even
increase it.

\begin{figure*}[htbp]
    \centering
    \includegraphics[width=15.5 cm]{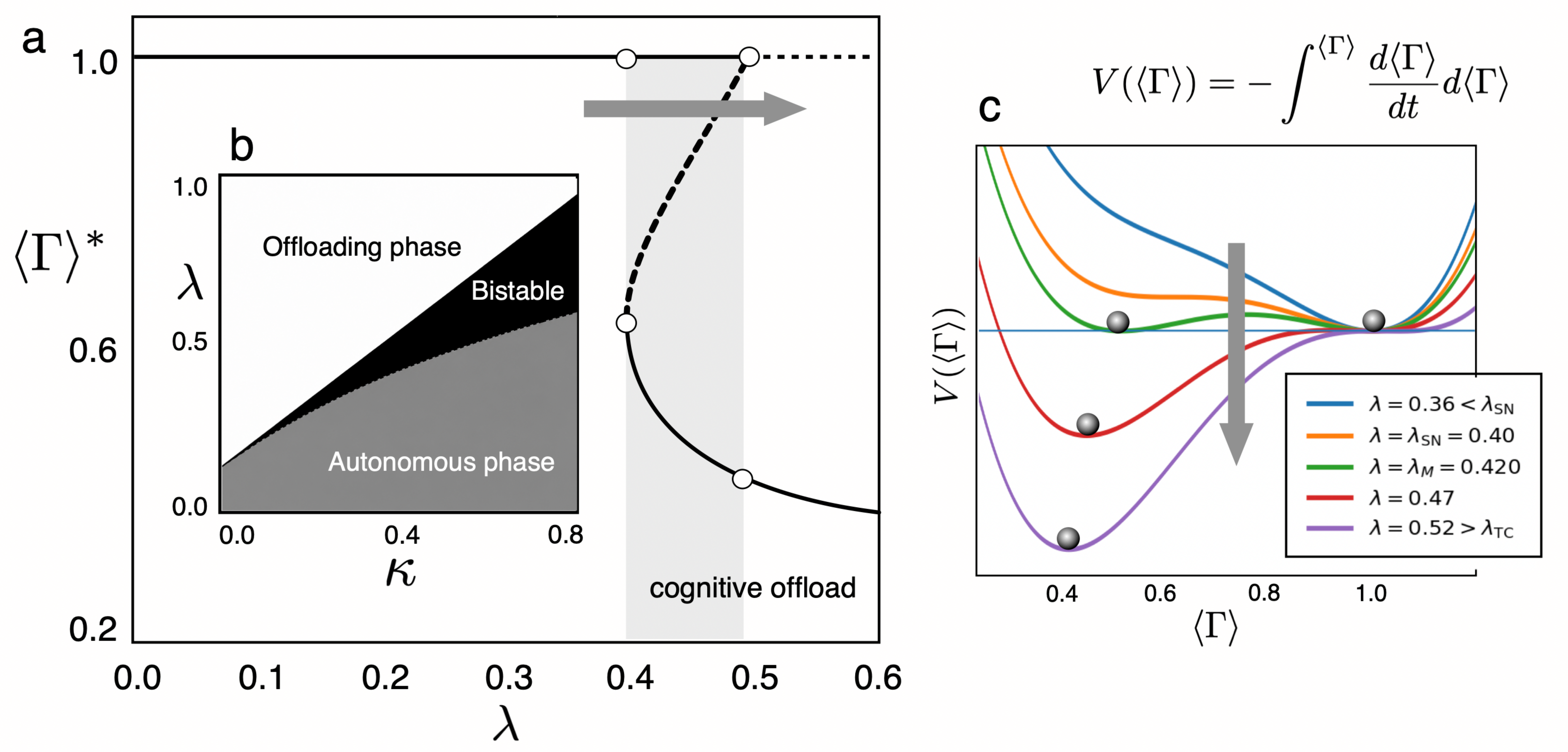}
\caption{\footnotesize
Cognitive offloading bifurcation, phase structure, and effective potential of the LLM propagation model. (a) Equilibrium average cognitive competence $\langle\Gamma\rangle^*$ as a function of the transmission parameter $\lambda$. The fully autonomous state remains stable up to the transcritical threshold $\lambda_{\rm TC}=\rho+\kappa=0.50$, whereas a stable offloading branch and an unstable branch appear at the saddle-node threshold $\lambda_{\rm SN}=2\sqrt{\kappa\rho}=0.40$. The shaded interval $\lambda_{\rm SN}<\lambda<\lambda_{\rm TC}$ marks the bistable region, where autonomous and offloaded states coexist, implying hysteresis and path dependence. (b) Phase diagram in the $(\kappa,\lambda)$ plane. The lower boundary, $\lambda_{\rm SN}=2\sqrt{\kappa\rho}$, and the upper boundary, $\lambda_{\rm TC}=\rho+\kappa$, delimit the bistable domain (black); below it the autonomous phase dominates (gray), whereas above it only the offloading phase is stable (white). (c) Effective potential $V(\langle\Gamma\rangle)$ for representative values of $\lambda$, following the gray arrow in (a) that crosses the bistable domain. The potential landscape shows how the single autonomous minimum is replaced by two competing minima in the bistable regime and finally by a single offloading minimum above $\lambda_{\rm TC}$. At $\lambda=\lambda_M$ (the so called Maxwell point), the two minima have equal depth. Same parameters as in Fig. \ref{fig:LLMbifurcations1}, with $\Gamma_u=1$, $\Gamma_c=0.5$, and $\Gamma_d=0.1$
}
    \label{fig:LLMbifurcations2}
\end{figure*}

Using the previous attractor states, the average cognitive competence becomes (see SM for a full derivation) 
\begin{equation}
\langle\Gamma\rangle^*
=g+(1-g)U^*,
\label{eq:gammaU}
\end{equation}
where
\begin{equation}
g=\frac{\Gamma_c\sigma+\Gamma_d\mu}{\mu+\sigma}
\end{equation}
is the mean effective CC of the coupled population. Thus,
$\langle\Gamma\rangle^*$ is an increasing function of $U^*$ and
inherits exactly the same saddle-node, bistable, and hysteretic structure
as the population dynamics (Fig. \ref{fig:LLMbifurcations2}a-b). In particular, when
$\kappa>\rho$, increasing $\lambda$ leaves the system on the fully
uncoupled state, $\langle\Gamma\rangle^*=1$, until
$\lambda_{\rm TC}=\rho+\kappa$. At this point the autonomous attractor loses
stability and the population moves to the coupled branch, for which
$U_-^*=\rho/\kappa$ at the transition. The corresponding drop in CC
is therefore
\begin{equation}
\Delta\Gamma_{\rm loss}
=(1-g)\left(1-\frac{\rho}{\kappa}\right).
\end{equation}
Conversely, when $\lambda$ is decreased from the coupled regime, recovery
does not occur at the same point. The system remains on the low-CC
branch until the saddle-node
$\lambda_{\rm SN}=2\sqrt{\kappa\rho}$ is reached, where
$U^*=\sqrt{\rho/\kappa}$. The subsequent return to the autonomous state
produces a CC increase
\begin{equation}
\Delta\Gamma_{\rm rec}
=(1-g)\left(1-\sqrt{\frac{\rho}{\kappa}}\right).
\end{equation}
cognitive competence therefore displays the same hysteresis as LLM
adoption: gradual changes in transmission pressure can generate an abrupt
loss of autonomous capacity, while reversing that loss requires a larger
reduction in $\lambda$.

The bifurcation structure itself is independent of this particular
choice of $\Gamma$; the values used here illustrate the consequences
of a regime in which increasing dependence is associated with reduced CC.
For the parameters used in Fig.~\ref{fig:LLMbifurcations1}, we obtain
$g=7/30\simeq0.233$. As $\lambda$ increases through
$\lambda_{\rm TC}=0.50$ (gray arrows), the equilibrium CC consequently falls
from $1$ to approximately $0.425$. In the reverse trajectory, the coupled
state persists to $\lambda_{\rm SN}=0.40$, where
$\langle\Gamma\rangle^*\simeq0.617$, before recovering to unity. The bifurcation therefore converts a smooth change in the pressure to adopt LLM-mediated cognition into a discontinuous and history-dependent change in population-level cognitive competence. Notice
also that $\mu$ and $\sigma$ do not alter the positions of the two tipping
points in this minimal model, but they determine $g$ and hence the cognitive
cost associated with occupying the coupled state. Importantly, the presence of this sharp transition has a key implication: 
crossing the bifurcation changes the attractor structure and can therefore produce an abrupt change in the equilibrium population state under quasistatic parameter variation. The minimal model does not, however, determine how rapidly the transition unfolds in real time.

An alternative visualization of this transition is provided by the effective
potential associated with the dynamics of $\langle\Gamma\rangle$. Under the
one-dimensional reduction described in the SM, the coupled fractions $C$ and
$D$ rapidly relax to their quasi-equilibrium ratio. The resulting dynamics can be
written as a gradient system,
\begin{equation}
\frac{d\langle\Gamma\rangle}{dt}
=-\frac{dV}{d\langle\Gamma\rangle},
\end{equation}
where the potential $V$ is obtained from: 
\begin{equation}
V(\langle\Gamma\rangle) = - \int^{\langle\Gamma\rangle} \left ( {d \Gamma \over dt}\right ) d\Gamma
\end{equation}
is a quartic function with two wells (see SM).
This representation preserves the equilibrium states and bifurcation
thresholds of the population model: minima of $V$ correspond to stable
population states, whereas the intervening maximum represents the unstable
branch separating their basins of attraction.

The evolution of this landscape as $\lambda$ increases is shown in Fig.~\ref{fig:LLMbifurcations2}c. The dynamics can be pictured as a marble
moving over the landscape until it settles in a valley. For low values of $\lambda$, the only valley lies at high 
$\langle\Gamma\rangle$, corresponding to strong cognitive autonomy. At $\lambda_{\rm SN}$, a second valley appears at
lower $\langle\Gamma\rangle$, representing a state dominated by cognitive offloading. Within the bistable interval, 
$\lambda_{\rm SN}<\lambda<\lambda_{\rm TC}$, either state can persist, depending on the system's history and on which
side of the intervening barrier it lies. As $\lambda$ increases, the low-$\langle\Gamma\rangle$ valley becomes 
progressively deeper. At   $\lambda_M\simeq0.420$ (the so called Maxwell point) for the parameters used here, the two 
valleys have equal depth; beyond it, the offloading state becomes increasingly favored. Nevertheless, the marble may 
remain trapped in the high-autonomy valley until this valley disappears at $\lambda_{\rm TC}$. It then rolls abruptly
toward the low-$\langle\Gamma\rangle$ minimum, producing a runaway transition: greater reliance on LLMs promotes 
further cognitive offloading, which in turn reinforces that reliance. In contrast, recovery of autonomy requires 
crossing the barrier or reducing $\lambda$ sufficiently so that the offloading valley disappears at $\lambda_{\rm SN}$.
The landscape thus provides an intuitive picture of tipping, runaway change, and hysteresis between autonomous and 
cognitively offloaded states.

\begingroup
\squeezetable
\begin{table*}[ht]
\centering
\setlength{\extrarowheight}{4pt}
\begin{ruledtabular}
\begin{tabular}{@{}lll@{}}

\textbf{Intervention} &
\textbf{Main mathematical effect} &
\textbf{Effect on tipping and cognitive state} \vspace{4pt}\\
\hline

\parbox[t]{5.0cm}{\raggedright\arraybackslash
Reduce propagation of substitutive/dependency-producing coupling} &
\parbox[t]{5.8cm}{\noindent\justifying
Decreases $\lambda$ by limiting automatic adoption and social or
institutional amplification of LLM use.} &
\parbox[t]{4.6cm}{\noindent\justifying
Direct. Prevents invasion for $\lambda<\lambda_{\rm TC}$; after lock-in,
recovery requires $\lambda<\lambda_{\rm SN}$.} \vspace{4pt}\\

\hline

\parbox[t]{5.0cm}{\raggedright\arraybackslash
Preserve autonomous alternatives and routes back to unaided cognition
} &
\parbox[t]{5.8cm}{\noindent\justifying
Increases $\rho$, the rate at which regular users return to uncoupled or
weakly coupled cognition. } &
\parbox[t]{4.6cm}{\noindent\justifying
Raises both thresholds and reduces hysteresis. Bistability disappears for
$\rho\geq\kappa$.} \vspace{4pt}\\

\hline

\parbox[t]{5.0cm}{\raggedright\arraybackslash
Strengthen collective autonomy} &
\parbox[t]{5.8cm}{\noindent\justifying
Increases $\kappa$, the strength of nonlinear collective reinforcement of
autonomous cognition.} &
\parbox[t]{4.6cm}{\noindent\justifying
Raises resistance to invasion, but for $\kappa>\rho$ also widens the
hysteretic interval.} \vspace{4pt}\\

\hline

\parbox[t]{5.0cm}{\raggedright\arraybackslash
Favor recovery over collective lock-in} &
\parbox[t]{5.8cm}{\noindent\justifying
Increases the ratio $\rho/\kappa$, strengthening individual routes back to
autonomous cognition relative to collective reinforcement.} &
\parbox[t]{4.6cm}{\noindent\justifying
Moves the system toward $\rho=\kappa$, where the saddle-node and
transcritical thresholds merge and the transition becomes continuous.}
\vspace{4pt}\\

\hline

\parbox[t]{5.0cm}{\raggedright\arraybackslash
Prevent progression to dependency} &
\parbox[t]{5.8cm}{\noindent\justifying
Decreases $\mu$, reducing transitions from regular use $C$ to persistent
dependency $D$.} &
\parbox[t]{4.6cm}{\noindent\justifying
No direct effect on $\lambda_{\rm SN}$ or $\lambda_{\rm TC}$.
Reduces $D^*/(C^*+D^*)$ and increases $\langle\Gamma\rangle$.}
\vspace{4pt}\\

\hline

\parbox[t]{5.0cm}{\raggedright\arraybackslash
Promote recovery from dependency} &
\parbox[t]{5.8cm}{\noindent\justifying
Increases $\sigma$, shifting users from persistent dependency $D$ back to
regular use $C$.} &
\parbox[t]{4.6cm}{\noindent\justifying
No direct effect on the bifurcation thresholds. Reduces dependency and
raises $\langle\Gamma\rangle$ at fixed $U^*$.} \vspace{4pt}\\

\end{tabular}
\end{ruledtabular}

\caption{
\textbf{Qualitative effects of cognitive immunization strategies.}
The parameters $\lambda$, $\rho$, and $\kappa$ modify the bifurcation
structure, with
$\lambda_{\rm SN}=2\sqrt{\kappa\rho}$ and
$\lambda_{\rm TC}=\rho+\kappa$.
For $\kappa>\rho$, these thresholds delimit the bistable regime.
By contrast, $\mu$ and $\sigma$ do not move the tipping points in this
minimal model, but determine the partition between regular and dependent
users and therefore the cognitive competence
$\langle\Gamma\rangle$.
}
\label{tab:interventions}

\end{table*}
\endgroup

\section{Cognitive immunization and intervention strategies}
The bifurcation structure provides a natural framework for considering
``cognitive immunization,'' understood here not as preventing contact with
LLMs, but as preserving resistance to harmful substitutive or
dependency-producing coupling while allowing beneficial forms of human-AI
integration. A first
important result is the asymmetry between prevention and reversal. If the
population is initially close to the uncoupled state $E_U$, invasion is
prevented provided
\begin{equation}
\lambda<\lambda_{\rm TC}=\rho+\kappa.
\end{equation}
Once the system has moved to the coupled attractor, however, restoring
$\lambda$ below $\lambda_{\rm TC}$ is insufficient whenever the system lies
inside the bistable regime. The coupled state persists until
\begin{equation}
\lambda<\lambda_{\rm SN}=2\sqrt{\kappa\rho}.
\end{equation}
Thus, after the tipping point has been crossed, reversal requires a larger
reduction in transmission pressure than would have been required to prevent
the transition in the first place. This prevention-reversal asymmetry is
the direct dynamical consequence of hysteresis and provides a simple
mechanism for technological lock-in.

Among the parameters controlling the bifurcation structure, $\rho$ has a
particularly clear protective role. Increasing $\rho$, the rate at which
regular users return to uncoupled or weakly coupled cognition, raises the
invasion threshold $\lambda_{\rm TC}$ while simultaneously moving the
system toward the boundary $\rho=\kappa$. At this boundary the physically
accessible saddle-node disappears; for $\rho\geq\kappa$, bistability is
lost and the transition becomes continuous. Interventions that make
autonomous cognition an accessible and recurrent state-for example,
through protected unaided tasks, deliberate periods of disengagement,
maintenance of non-LLM skills, or attractive non-LLM alternatives-can
therefore modify not only the position of the tipping point but the
qualitative form of the transition itself.

The effect of $\kappa$ is more subtle. Increasing $\kappa$ strengthens the collective reinforcement of autonomous cognition and raises $\lambda_{\rm TC}$, making invasion more difficult when the population is predominantly uncoupled. At the same time, for $\kappa>\rho$, it also increases the separation between the two thresholds, $\Delta\lambda =\lambda_{\rm TC}-\lambda_{\rm SN}$ (Eq.\ \ref{eq:dLambda}), thereby enlarging the hysteretic region. Strong collective protection can therefore stabilize autonomy while $U$ is high, but may also increase path dependence once the autonomous fraction has been substantially depleted. The relevant intervention is consequently not simply to maximize $\kappa$, but to reinforce collective autonomy together with sufficiently strong return processes $\rho$ so that protection does not come at the cost of a broad hysteretic regime.

The parameters $\mu$ and $\sigma$ play a different role. They do not shift
$\lambda_{\rm SN}$ or $\lambda_{\rm TC}$, but control
the composition of the coupled population. At equilibrium, we have
\begin{equation}
\frac{D^*}{C^*+D^*}
=\frac{\mu}{\mu+\sigma},
\end{equation}
so decreasing $\mu$ or increasing $\sigma$ reduces the fraction of users in
the dependent state without necessarily reducing overall LLM adoption.
These parameters therefore describe a second class of interventions aimed
not at preventing coupling itself, but at limiting its transition toward
persistent cognitive substitution. Examples include metacognitive
training, verification requirements, periodic unaided practice, task
designs that preserve active reasoning, and mechanisms that facilitate
recovery from dependency.

The model thus distinguishes two complementary intervention levels.
Changes in $\lambda$, $\rho$, and $\kappa$ reshape the population-level
tipping landscape and determine whether bistability and hysteresis are
possible, whereas changes in $\mu$ and $\sigma$ primarily determine the
cognitive burden associated with the coupled state. In this sense,
immunization can act either by preventing a collective transition or by
reducing the probability that ordinary LLM use develops into persistent
dependency.

Thus immunization (as defined here) need not imply low overall LLM use. It can instead
consist of shaping the coupling so that high adoption remains compatible with
verification, active reasoning, autonomous alternatives, and recovery from
substitutive dependence.

\section{Discussion}

In 1960, Joseph Licklider anticipated a future in which the relation between humans 
and computers would move beyond simple tool use \cite{licklider1960man}:

\begin{quote}
``Man-computer symbiosis is probably not the ultimate paradigm for complex 
technological systems. It seems entirely possible that, in due course, electronic 
or chemical machines will outdo the human brain in most of the functions we now 
consider exclusively within its province.''
\end{quote}

More than six decades later, LLMs made this quote especially concrete. Their 
importance lies not only in what they can do, but in the new forms of coupling they 
create between human and artificial cognition. At one extreme, AI could become an 
``exocortex'': an external extension of cognition capable of searching large bodies of 
knowledge while leaving interpretation and high-level judgment to the human researcher 
\cite{yager2024exocortex}. Such systems could greatly accelerate 
the accumulation and transmission of cultural and scientific knowledge. More 
generally, humans have always externalized parts of cognition into language, writing, 
institutions, and technology. This externalization has been the engine of rapid and accelerated 
change driven by culture \cite{muthukrishna2018cultural,perreault2026cultural}. AI 
represents a new step in this process, one in which 
external cognitive machinery becomes increasingly active and agent-like 
\cite{ruffini2026pattern}. As suggested in previous studies, this might be 
a new class of {\it synthetic} evolutionary transition \cite{sole2016synthetic,rainey2023major,rainey2025could,muller2026evolvable}.

Our model suggests that the consequences of this transition cannot be 
understood from individual LLM use alone. Two social processes interact. First, new 
ways of using LLMs spread because people learn from other people, because workplaces 
and schools adopt them, and because successful practices are copied. In this sense, 
adoption has a contagion-like component. Second, autonomous cognition is itself 
socially sustained. Education and institutions do 
more than transmit information: they create environments in which reading, writing, 
reasoning, verification, and independent problem solving are repeatedly practiced 
and rewarded. When these practices are common, they reinforce each other.

The interaction between these two processes creates the central feedback in our 
model. As cognitive delegation becomes more widespread, the social environment that 
supports autonomous reasoning can weaken; as that environment weakens, delegation 
becomes still easier and more attractive. Therefore, a gradual increase in LLM adoption can
 produce a disproportionate collective response. Beyond a critical point, 
the loss of autonomy becomes self-reinforcing and the population can move rapidly 
toward a state of much stronger cognitive offloading and lower cognitive competence. The important result is not that such a runaway must occur, but that it 
\emph{can} occur under plausible forms of social learning and cooperative reinforcement. 
Moreover, once such a state has become established, simply returning conditions to 
where they were before the transition may not be sufficient to restore the previous 
state. Prevention can therefore be considerably easier than reversal.

This possibility should not obscure the substantial benefits of LLMs. The distinction 
is not between using and not using AI, but between forms of coupling that extend human 
competence and those that replace the cognitive operations through which competence 
is maintained. The evidence already points to both possibilities. 
Recent experiments also show that the direction of these effects strongly depends
 on the architecture of the human-AI coupling. In a randomized study, a
guided ``think first, ChatGPT later'' protocol produced higher subsequent independent creativity than unrestricted use of the ChatGPT \cite{wong2026think}. Moreover, generative AI can 
increase productivity and provide powerful cognitive assistance, but knowledge 
workers also report a reduced effort to think critically when confidence in AI is high 
\cite{lee2025critical}. In education, unrestricted access to generative AI can improve 
performance while the tool is available, yet reduce subsequent unaided performance, 
while pedagogically constrained AI can substantially mitigate this effect 
\cite{bastani2025guardrails,huettig2024can}. Related results indicate poorer comprehension or 
retention when students rely on LLMs without engaging in complementary cognitive 
activities such as note-taking \cite{kreijkes2026llm}. In programming, novice users 
can also struggle to understand and critically evaluate AI-generated code, creating 
conditions for automation bias and superficial competence \cite{zi2025code}. 

These 
findings reinforce the distinction between AI as \emph{scaffolding} and AI as 
\emph{substitution}. This distinction also changes how we should think about ``cognitive immunization.'' 
At the population level, immunization does not mean preventing contact with AI. It 
means preserving the practices and institutions that keep human cognition active: 
unaided problem solving, verification, critical discussion, periods of deliberate 
disengagement, maintenance of non-AI skills, and educational designs in which the 
model supports rather than completes the cognitive task. However, once persistent dependency 
has developed, the problem can no longer be addressed by educational design 
alone. Emerging work on problematic LLM use points to psychological and behavioral 
mechanisms related to loss of control, emotional regulation, cognitive biases, and 
habitual reliance, suggesting that behavioral self-regulation and, in more severe 
cases, psychological interventions such as cognitive-behavioral therapy may become 
relevant \cite{liao2026problematic}. Thus, the analogy with immunization extends naturally from prevention at the cultural and institutional level to recovery 
at the individual level.
  
Several extensions could bring this minimal model closer to the complexity of real human-LLM interactions. These include replacing the population-level description with heterogeneous interacting agents \cite{krakauer2026competitive}, allowing cognitive autonomy to vary continuously rather than dividing users into sharply separated classes, and introducing feedbacks through which cognitive change modifies subsequent adoption and dependence. Such a formulation would also soften the boundaries between humans, LLMs, and increasingly hybrid human-machine systems \cite{sole2026cognition}, which may occupy a broad continuum of cognitive organizations rather than discrete categories. More generally, if agents themselves can be understood as persistent patterns of information and activity distributed on biological, artificial, or hybrid substrates \cite{levin2022technological, FIELDS2025256, ruffini2026pattern}, the relevant evolving entities may not always be conventional individuals. What spreads could instead be patterns of cognitive organization, including perspectives that shape how agents represent the world, themselves, and possible actions. LLMs would then be more than passive tools for cognitive offloading: by participating in these distributed patterns, they could progressively reshape the cognitive structures through which users interpret and act on the world. From this perspective, the ``cognitive virus'' is a persistent technological lineage embedded in a larger ecology
of cultural replicators, human host states, institutions, and technical
infrastructure. 

A further limitation is that human behavior is not yet explicitly coupled 
to the spreading process. Such feedbacks can qualitatively reshape epidemic 
thresholds~\cite{fenichel2011adaptive,bootsma2007effect,funk2010modelling,
funk2009spread,bosetti2020heterogeneity}, and can even generate higher-order 
critical phenomena~\cite{granell2013dynamical,lima2015disease,
maniscalco2025critical}. A useful conceptual framework for extending this 
picture is provided by the emerging field of \textit{Diverse Intelligence}, 
which treats intelligence and agency not as properties restricted to brains or 
particular biological substrates, but as graded capacities of systems operating 
across multiple spatial, temporal, and organizational scales 
\cite{ChisCiureLevin2025,Levin2025Artificial}. In particular, this perspective 
emphasizes the ability of biological, artificial, and hybrid agents \cite{sole2026cognition} to navigate 
problem spaces and maintain goal-directed organization, providing a natural 
language to describe coupled human-AI systems in which boundaries and 
degrees of agency may themselves change over time. In addition, 
multilayer-network theory offers a formal framework for representing the 
interacting behavioral, technological, and social processes through which such 
feedbacks propagate~\cite{de2016physics,de2023more}. Together, these approaches 
suggest a systematic route beyond the present mean-field description.

Future models could therefore treat autonomy, dependence, and agency as 
continuous and coevolving properties of distributed human-machine patterns, 
rather than as fixed compartments, and ask under what conditions transient 
interactions become self-maintaining forms of cognitive organization.

\begin{acknowledgments}
R.S. has been supported by an AGAUR 2021 SGR 0075 grant, by AEI-PID2023-152129NB-I00 grant funded by MICIU/AEI/10.13039/501100011033 and ERDF, EU, and the Santa Fe Institute. S.F.E. has been supported by grants PID2025-169610NB-I00 funded by MCIU/AEI/10.13039/501100011033 and by “ERDF a way of making Europe”, CIPROM/2022/59 funded by Generalitat Valenciana, and the Santa Fe Institute. L.F.S. has been supported by the Occident Foundation (grant FJSCNB-2022-12-B) and by the Spanish State Research Agency, AEI, through grant PID2023-153225NA-I00 funded by MICIU/AEI/10.13039/501100011033 and ERDF, EU. L.F.S. conducted this research while visiting the Okinawa Institute of Science and Technology (OIST) through the Theoretical Sciences Visiting Program (TSVP).
\end{acknowledgments}

\renewcommand{\bibsection}{\section*{References}}

%

\end{document}